\documentclass[12pt]{iopart}

\usepackage{graphicx}
\usepackage{graphics}
\usepackage{multirow}
\usepackage{hyperref}
\hypersetup{
    colorlinks=true,
    linkcolor=blue,
    filecolor=blue,      
    urlcolor=blue,
    citecolor=blue
}

\newcommand{\eqref}[1]{(\ref{#1})}

\begin{document}

\title[Synchrotron images in Alcator C-Mod]{Spatiotemporal evolution of runaway electrons from synchrotron images in Alcator C-Mod}

\author{R.A. Tinguely$^1$\footnote{Author to whom correspondence should be addressed: rating@mit.edu}, R.S. Granetz$^1$, M. Hoppe$^2$, and O. Embr\'{e}us$^2$}

\address{$^1$ Plasma Science and Fusion Center, Massachusetts Institute of Technology, Cambridge, MA, USA \\
$^2$ Department of Physics, Chalmers University of Technology, G\"{o}teborg, Sweden}

\begin{abstract}

In the Alcator C-Mod tokamak, relativistic runaway electron (RE) generation can occur during the flattop current phase of low density, diverted plasma discharges. Due to the high toroidal magnetic field ($B_0$~=~5.4~T), RE synchrotron radiation is measured by a wide-view camera in the visible wavelength range ($\lambda \approx$~400-900~nm). In this paper, a statistical analysis of over one thousand camera images is performed to investigate the plasma conditions under which synchrotron emission is observed in C-Mod. In addition, the spatiotemporal evolution of REs during one particular discharge is explored in detail via a thorough analysis of the distortion-corrected synchrotron images. To accurately predict RE energies, the kinetic solver CODE [Landreman \emph{et al} 2014 \emph{Comput. Phys. Commun.} \textbf{185} 847-855] is used to evolve the electron momentum-space distribution at six locations throughout the plasma: the magnetic axis and flux surfaces $q$~=~1, 4/3, 3/2, 2, and 3. These results, along with the experimentally-measured magnetic topology and camera geometry, are input into the synthetic diagnostic SOFT [Hoppe \emph{et al} 2018 \emph{Nucl. Fusion} \textbf{58} 026032] to simulate synchrotron emission and detection. Interesting spatial structure near the surface $q$~=~2 is found to coincide with the onset of a locked mode and increased MHD activity. Furthermore, the RE density profile evolution is fit by comparing experimental to synthetic images, providing important insight into RE spatiotemporal dynamics.

\end{abstract}


%%%%%%%%%%%%%%%%%%%%%%
%%%%%%%%%%%%%%%%%%%%%%
% Introduction 
%%%%%%%%%%%%%%%%%%%%%%
%%%%%%%%%%%%%%%%%%%%%%

\section{Introduction}\label{sec:introduction}

In a tokamak plasma, a toroidal loop voltage is externally applied to accelerate electrons and drive a plasma current, $I_P$. These electrons experience collisions with other plasma particles with a characteristic time that increases with their speed, $\tau_{coll} \propto v^3$. Thus, a sufficiently strong electric field can overcome collisional friction and continuously accelerate electrons to relativistic speeds. These so-called ``runaway electrons'' (REs) are often generated in one of three phases of the plasma discharge: during (i) plasma start-up, when $I_P$ steadily increases; (ii) the steady-state ``flattop $I_P$'' portion, if the electron density is sufficiently low; or (iii) plasma disruptions, when the thermal and magnetic energies rapidly dissipate. REs have been experimentally measured to attain energies of tens of MeV and carry currents $>50\%$ of $I_P$ \cite{martin-solis2006,plyusnin2006,plyusnin2018}. Thus, the impact of REs with the tokamak first wall can cause serious damage to plasma-facing components, and their formation --- though a fascinating plasma phenomenon --- should be avoided (or mitigated) in future fusion devices. 

Traditionally, the primary diagnostic for REs is measurement of hard x-ray (HXR) emission and spectra; however, this bremsstrahlung typically results when REs lose confinement and hit the vessel wall. To study REs ``in-flight,'' RE synchrotron emission can be used. This relativistic cyclotron emission arises primarily from RE gyromotion in the background toroidal magnetic field, $B$. Since synchrotron radiation is directed along each RE's velocity vector, it is only seen from the counter-$I_P$ direction and thus on one side of the tokamak (see figure~\ref{fig:wide2}). Depending on $B$ and the RE energy and pitch (ratio of velocities perpendicular and parallel to the magnetic field, $v_\perp/v_\parallel$), synchrotron emission can be measured in the infrared and even visible wavelength ranges \cite{schwinger1949}. While synchrotron spectra can provide insight into the energy distribution of REs, camera images of synchrotron emission capture the spatiotemporal evolution of REs throughout the plasma. Table~\ref{tab:experiments} gives an overview of previous RE studies in which REs were diagnosed using synchrotron images. Experiments are discriminated by their RE generation phase --- during plasma start-up (S), flattop $I_P$ (F), or disruption (D) --- and camera type --- visible (V) or infrared (IR). In addition, different studies analyzed synchrotron images in various ways. Here, the ``spatial dimensionality'' is reported: In some studies, only 0D time evolutions of synchrotron intensity (e.g. the total number of photon counts within the detected image) were analyzed; in others, 1D radial profiles (e.g. vertical integrations of a horizontal camera ``slit'') were used, often to explore radial diffusion. Two-dimensional data have been used to study the spatial properties of the RE beam: height and width, often related to the pitch angle; shape, like crescents or hollow rings; or feature-mapping to flux surfaces and drift orbits. Note that many works utilizing 2D information also often perform 0D and 1D analyses. Some efforts have been made to go beyond the identification of spatial features and to analyze the synchrotron intensity distribution throughout the image, as does the present study; these are bolded in table~\ref{tab:experiments}.  

\begin{table}[h]
    \centering
    \caption{An overview of RE synchrotron image measurements organized by device; RE generation during start-up (S), plasma current flattop (F), or disruption (D) phases; visible (V) or infrared (IR) camera type; and highest ``spatial dimensionality'' of the analysis (as described in section~\ref{sec:introduction}). \textbf{Bold} entries indicate analysis of the intensity distribution beyond spatial features (e.g. width, shape, etc.).}
    \label{tab:experiments}
    \begin{tabular}{| r | c c c c |}
        \hline
        Device & RE generation & Camera type & Spatial analysis & References \\
        \hline
        Alcator C-Mod & F & V & \textbf{2D} & \cite{hoppe2018} \\
        \hline
        COMPASS & F & IR & 2D &\cite{vlainic2015} \\
        \hline
        \multirow{5}{*}{DIII-D} & \multirow{3}{*}{F} & \multirow{3}{*}{V} & 0D & \cite{paz-soldan2017,paz-soldan2018} \\
         & & & 2D & \cite{paz-soldan2014} \\
         & & & \textbf{2D} & \cite{hoppe2018d3d}  \\
         \cline{2-5}
         & \multirow{2}{*}{D} & V+IR & \multirow{2}{*}{2D} & \cite{hollmann2015} \\
         & & V & & \cite{yu2013} \\
        \hline
        \multirow{2}{*}{EAST} & S+F & \multirow{2}{*}{V} & \multirow{2}{*}{2D} & \cite{zhou2013,zhou2014} \\
         & F & & & \cite{shi2010}\\
        \hline
        FTU & F & V+IR & 0D & \cite{esposito2017} \\
        \hline
        HL-2A & D & V & 2D & \cite{zhang2012} \\
        \hline
        HT-7 & F & IR & 2D & \cite{chen2006} \\
        \hline
        J-TEXT & F+D & IR & 2D & \cite{tong2016}  \\
        \hline
        KSTAR & S & IR & 2D & \cite{england2013} \\
        \hline
        \multirow{6}{*}{TEXTOR} & \multirow{3}{*}{D} & \multirow{3}{*}{IR} & 0D & \cite{bozhenkov2008} \\
         & & & 1D & \cite{wongrach2014} \\
         & & & 2D & \cite{wongrach2015}  \\
         \cline{2-3}\cline{4-5}
         & \multirow{3}{*}{F} & \multirow{3}{*}{IR} & 0D & \cite{entrop2000} \\
         & & & 1D & \cite{entrop1998,kudyakov2012} \\
         & & & 2D & \cite{finken1990,jaspers1994,entrop1999} \\
        \hline
    \end{tabular}
\end{table}

This work reports the analysis of 2D camera images of visible synchrotron emission from REs generated during the flattop phase of low density, diverted plasma discharges in the Alcator C-Mod tokamak. To the authors' knowledge, this is the most complete study of synchrotron images performed to-date. Novel to this work is the combination of the following: (i) Experimentally-measured spatial profiles of plasma parameters are used to simulate the time evolution of the RE momentum-space distribution \emph{throughout} the plasma \cite{landreman2014,stahl2016}; past studies typically calculate single particle momenta from plasma parameters only on-axis. (ii) The new synthetic diagnostic SOFT \cite{hoppe2018} is used to model the synchrotron intensity pattern detected by a camera, given experimentally-measured magnetic and detector geometries; most previous works do not account for these necessary geometric effects. (iii) Comparisons of the full 2D intensity distributions of synthetic and experimental images allow diagnosis of the time-evolving RE density profile; only recently have synchrotron image analyses moved beyond spatial feature identification.

Note that a similar procedure, as described above, was also used in \cite{tinguely2018} to study the time evolution and magnetic field dependence of RE synchrotron \emph{spectra}. However, in that study, spectral measurements were volume-integrated within each spectrometer's field-of-view, providing less spatial information than a camera. Consequently, a test particle approach for RE density and momentum evolution was sufficient to match experimental observations. As will be discussed in section~\ref{sec:TPM}, the full momentum-space distribution is required to reproduce experimental synchrotron images.

The organization of content is as follows: Section~\ref{sec:experimentation} details the experiment and setup. In section~\ref{sec:aggregate}, a statistical analysis is presented of aggregate data from many RE discharges with and without observed synchrotron emission. Section~\ref{sec:spatiotemporal} explores the spatiotemporal evolution of REs in one particular discharge, incorporating the aforementioned improved analysis techniques. Finally, a discussion and summary of results are given in section~\ref{sec:summary}.

%%%%%%%%%%%%%%%%%%%%%%
%%%%%%%%%%%%%%%%%%%%%%
% Experimentation
%%%%%%%%%%%%%%%%%%%%%%
%%%%%%%%%%%%%%%%%%%%%%

\section{Experiment and setup}\label{sec:experimentation}

\begin{figure}[h]
    \centering
    \includegraphics[width=\textwidth]{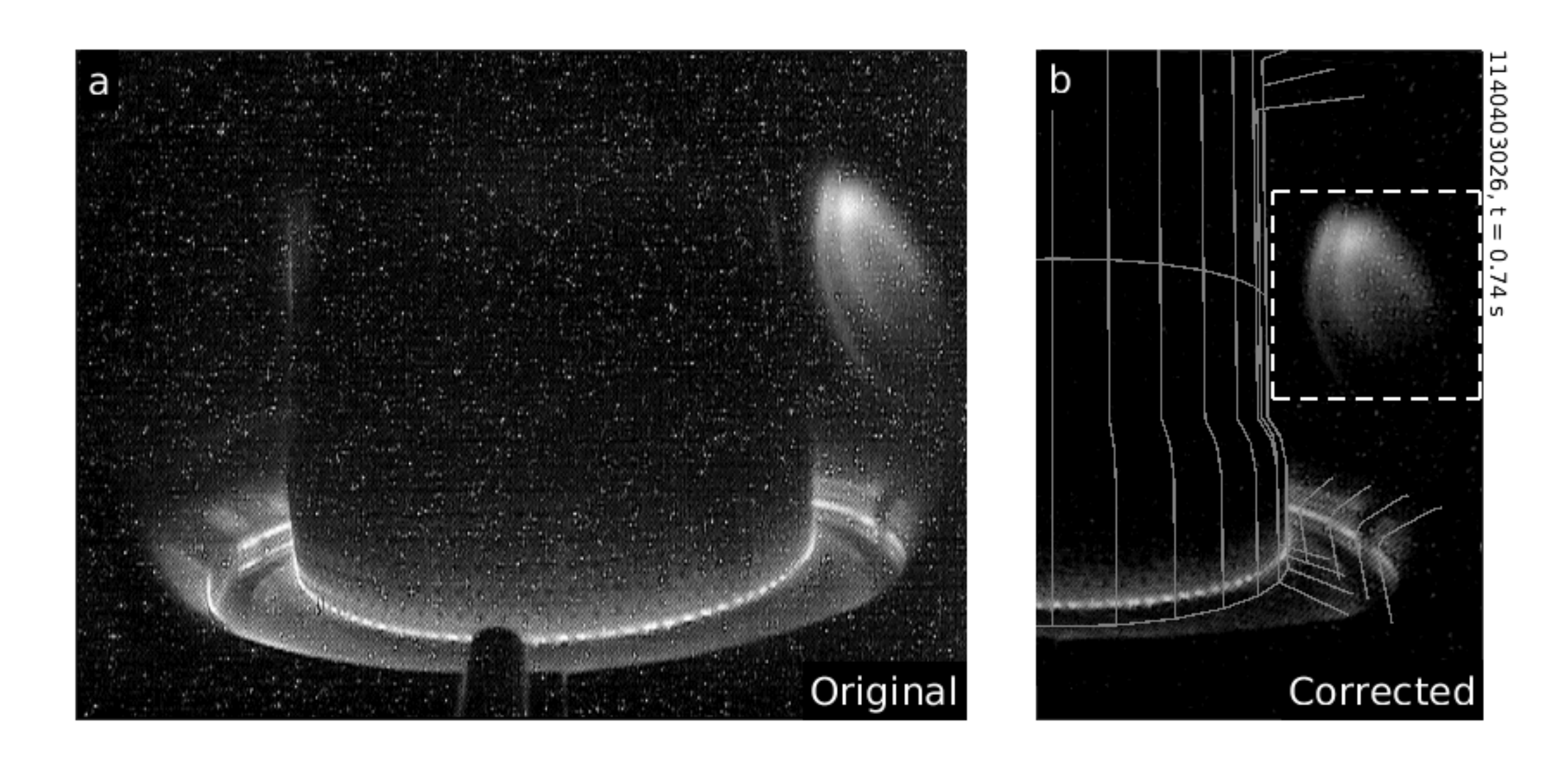}%{figures/WIDE2_originalFull_corrected_1140403026_t0740ms_180803.png}
    \caption{The (a) original camera image and (b) right half of the corrected image after distortion-correction, vertical-alignment, background-subtraction, and HXR-removal. A 2D projection of the 3D vacuum vessel geometry overlays the image. Intensity data within the dashed box are analyzed. ($t$~=~0.74~s)}
    \label{fig:wide2}
\end{figure}

The Alcator C-Mod tokamak is a high-field, compact fusion device with magnetic field on-axis ranging from $B_0$~=~2-8~T and major and minor radii of $R_0$~=~0.68~m and $a$~=~0.22~m, respectively. A wide-angle visible and near-IR camera, with wavelength range $\lambda \approx$~400-900~nm, is used for general monitoring of the vacuum vessel during the plasma discharge. The camera is located $\sim$21~cm below the midplane with a near radially-inward view, capturing images at $\sim$60 frames per second. See table~\ref{tab:wide2} for additional camera specifications. Because the peak of the synchrotron power spectrum shifts toward shorter wavelengths with increasing magnetic field strength (assuming fixed RE energy and pitch) \cite{schwinger1949}, synchrotron radiation is often measured by the camera in the visible-NIR wavelength range at C-Mod's operational field, $B_0$~=~5.4~T. An example of raw camera data is shown in figure~\ref{fig:wide2}a. The camera measures intensity only (i.e. black-and-white, not in color), so no spectral information is obtained. All other figures in this paper are false-colored to better highlight the intensity distribution. Note that light dominates on the right side of the image, indicating that this is in fact synchrotron emission. The white speckles on the image are the result of HXR radiation impacting the camera. In addition, the camera auto-gain is off; while this means that sometimes pixel saturation occurs, it also allows frame-by-frame comparisons of pixel intensity. 

\begin{table}
	\centering
	\caption{Specifications of the wide-view camera in Alcator C-Mod.}
	\label{tab:wide2}
	\begin{tabular}{c c}
		\hline
		Specification & Value \\
		\hline
		Aperture diameter & 7~mm \\
		Pixel dimensions & $640 \times 480$ \\
		Spectral range & $\sim$400-900~nm \\
		Major radial position & 106.9~cm \\
		Vertical position & -20.7~cm \\
		Frame rate & 59.94 fps\\
		Tilt (roll) & 3.9$^\circ$ \\
		Yaw &  1.9$^\circ$ \\
		Inclination (pitch) & 1.7$^\circ$ \\
		Total viewing angle & 86.8$^\circ$ \\
		\hline
	\end{tabular}
\end{table}

A fish-eye lens on the camera causes barrel distortion of the images. Therefore, an in-vessel calibration was performed to map raw pixel data to a rectilinear detector plane. An absolute calibration, mapping pixel intensity to Watts, was not performed, however. The tilt angle was assessed from the images by aligning the edge of the tokamak inner wall with the vertical axis. The right half of the corrected image is shown in figure~\ref{fig:wide2}b. Note how the vertical extent of the distortion-corrected synchrotron spot has been shortened significantly. This emphasizes the importance of camera calibration before synchrotron image processing and analysis. Background visible plasma emission, averaged from the co-$I_P$ direction (left side of figure~\ref{fig:wide2}a), has been subtracted in figure~\ref{fig:wide2}b, and most HXR speckles have been filtered out. A 2D projection of the 3D vacuum vessel geometry overlays the corrected image and matches features quite well. Vertical positions of the midplane ($Z$~=~0) and lower inboard divertor ``corner'' ($Z\approx-48$~cm) are also indicated in figure~\ref{fig:wide2}b. Because the camera is below the midplane, the synchrotron spot has a unique parabolic shape, different from the crescents, ellipses, and hollow rings seen in other tokamaks. The dashed box outlines the subset of pixels ($150 \times 150$) within which synchrotron emission is primarily observed; this region will be the focus of the following analyses. 

\begin{figure}[h]
    \centering
    \includegraphics[width=\textwidth]{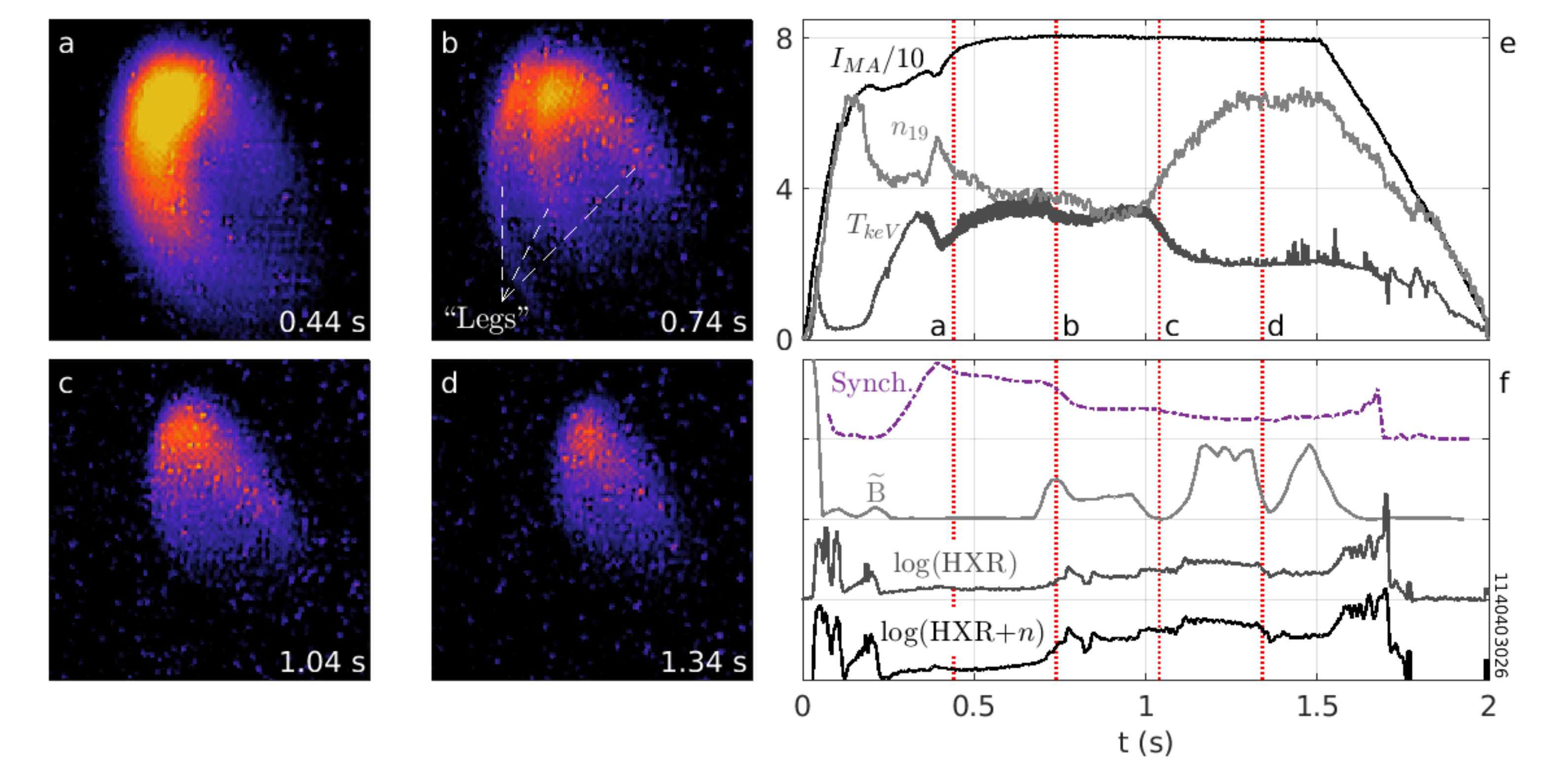}%{figures/180822/plasmaParameters_WIDE2_1140403026_t0440-0740-1040-1340ms_180822_v3.pdf}
    \caption{Corrected experimental images (false-colored) at four times: (a) 0.44, (b) 0.74, (c) 1.04, and (d) 1.34~s. The observed synchrotron spot ``legs'' are indicated by dashed lines in (b). Plasma parameters in (e) are plasma current (MA/10), line-integrated density (10$^{19}$ m$^{-3}$), and central temperature (keV); in (f) are signals (a.u.) of summed synchrotron intensity in each frame (dot-dashed), locked-mode amplitude $\widetilde{B}$, HXR radiation, and HXR+photoneutrons, each with a different vertical axis offset. Times (a)-(d) are marked by vertical dotted lines in (e)-(f).}
    \label{fig:params}
\end{figure}

During disruptions of diverted plasmas in C-Mod, magnetic flux surfaces rapidly become stochastic \cite{izzo2011}; therefore, high energy ($>$10~MeV) post-disruption REs are not observed as they quickly lose confinement \cite{marmar2009}. Instead, relativistic REs can be generated during plasma start-up and low-density steady-state discharges. Plasma parameters for a typical flattop RE discharge are shown in figure~\ref{fig:params}e. Camera images of the synchrotron spot (within the $150\times150$ pixel box of figure~\ref{fig:wide2}b) are shown at four times, indicated by the vertical dotted lines in figure~\ref{fig:params}e-f: $t = $~0.44, 0.74, 1.04, and 1.34~s. Initially, the plasma density, $n$, decreases in time, reducing collisional friction and encouraging RE growth. The measured synchrotron intensity (summed within each frame) is bright at $t\approx$~0.4~s, as seen in figure~\ref{fig:params}f; in fact, the synchrotron emission actually saturates the camera, as seen in figure~\ref{fig:params}a. 

At $t \approx$~0.7~s, the plasma rotation slows as a locked mode begins. This is determined by both a partial reduction of sawteeth in the temperature evolution, $T$, as well as magnetic fluctuations, $\widetilde{B}$, measured by Mirnov coils within the first wall.\footnote{These magnetic fluctuations actually correspond to a high frequency ($\sim$40-60~kHz) signal which is found to be correlated with locked modes in C-Mod.} There is a reduction in total synchrotron intensity at this time, and the synchrotron spot develops interesting spatial structure: As seen in figure~\ref{fig:params}b, there appear to be three ``legs'' to the synchrotron spot, i.e. distinct ``inner'' and ``outer'' legs on the high-field (left) side of the image. During the period of $\widetilde{B}$ fluctuations from $t\approx$~0.7-1.0~s, measured HXR and photoneutron signals also increase, indicating loss of REs to the first wall.

At $t$~=~1~s, the plasma density is increased to suppress RE growth. In response, the synchrotron spot decreases both in intensity and size, as seen in figure~\ref{fig:params}c. Around the same time, sawteeth disappear, suggesting a fully-locked plasma. The amplitude of $\widetilde{B}$ fluctuations increases as the RE beam continues to decrease in size and intensity (see figure~\ref{fig:params}d). Finally, at the end of the discharge, during the $I_P$ ramp-down, there is a sudden flash of synchrotron light and corresponding spikes in HXR signals, indicating the final loss of RE confinement.

\begin{figure}[h]
    \centering
    \includegraphics[width=0.65\textwidth]{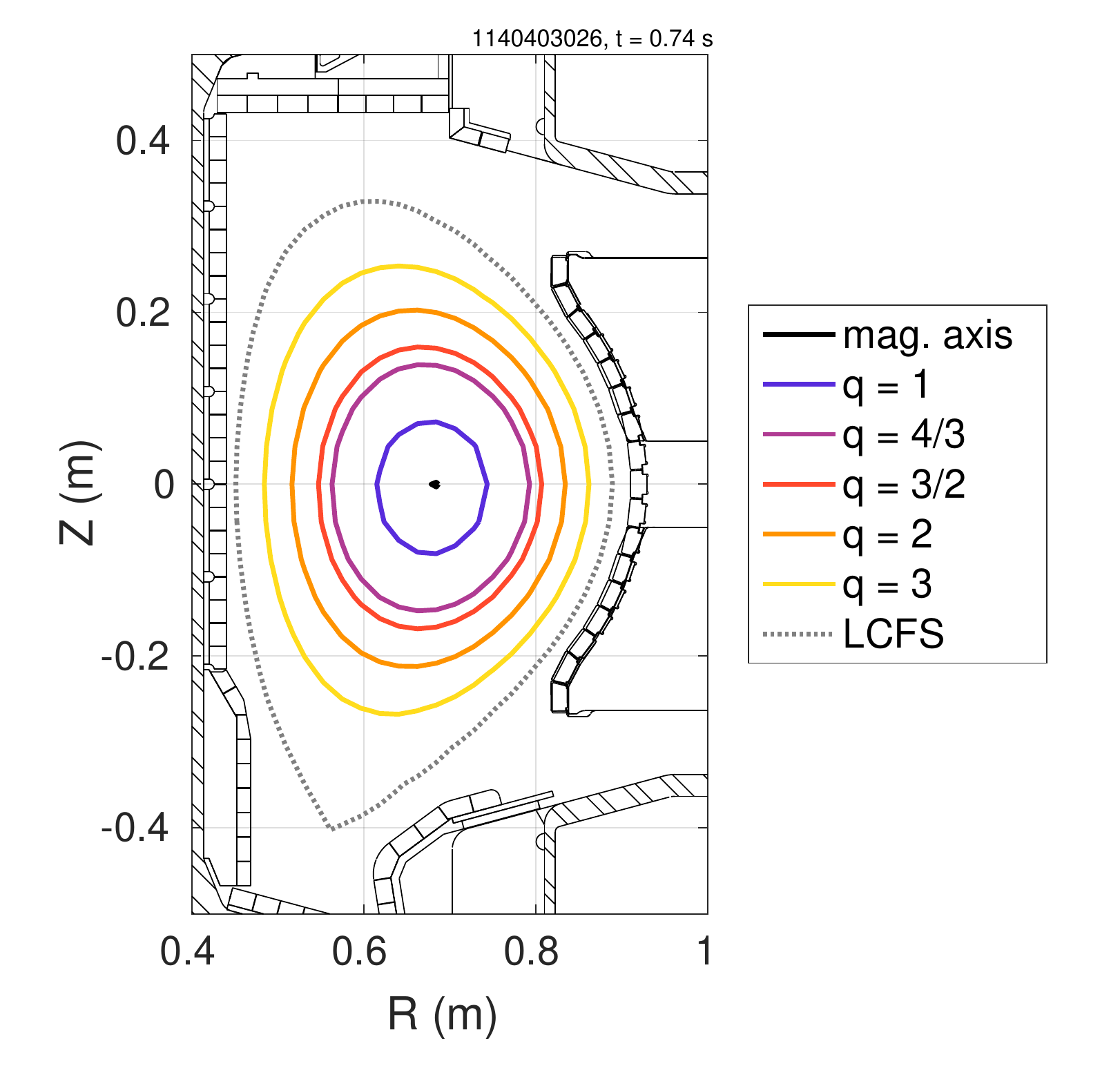}%{figures/180924/EFITContours_1140403026_t0740ms_180924.pdf}
    \caption{Overlaying a poloidal cross-section of the C-Mod vacuum vessel are poloidal flux contours from EFIT \cite{lao1985}: the magnetic axis (circle), rational surfaces $q$~=~1, 4/3, 3/2, 2, and 3 (solid), and the last closed flux surface (LCFS, dotted). ($q_{95} \approx 4.5$, $t$~=~0.74~s)}
    \label{fig:EFIT}
\end{figure}

Poloidal flux contours, as calculated by EFIT \cite{lao1985}, are shown for one time, $t$~=~0.74~s, in figure~\ref{fig:EFIT}. Note that this is the same time as the camera image in figure~\ref{fig:wide2}. Specifically highlighted are the magnetic axis, rational flux surfaces $q$~=~1, 4/3, 3/2, 2, and 3, as well as the last closed flux surface. It is important to note that all discharges analyzed in this work are elongated and diverted. In addition, EFIT reconstructions of the magnetic flux were constrained only on-axis, i.e. a bound was set around $q_{axis} \approx 1$ due to observations of sawtoothing. However, similar C-Mod plasma discharges, from different experimental runs, were able to perform more accurate magnetic reconstructions using data from the Motional Stark Effect (MSE) diagnostic. The MSE-constrained $q$-profiles closely match those used in this work, thus verifying the following analyses.

%%%%%%%%%%%%%%%%%%%%%%
%%%%%%%%%%%%%%%%%%%%%%
% Aggregate analysis
%%%%%%%%%%%%%%%%%%%%%%
%%%%%%%%%%%%%%%%%%%%%%

\section{Aggregate analysis}\label{sec:aggregate}

The original purpose of these C-Mod RE experiments --- from which the synchrotron images were obtained --- was to evaluate the critical electric field for RE generation and suppression, as reported in \cite{granetz2014}. However, in that study, periods of RE growth and decay were deduced from HXR signals, not synchrotron emission, since low energy REs ($<$~10~MeV) do not radiate in the visible wavelength range and therefore would not be detected by the camera. Here, the analysis of \cite{granetz2014} is extended to investigate the plasma conditions under which visible synchrotron emission was or was not observed. Many plasma discharges were reproduced throughout the experimental run (C-Mod \#1140403), varying such parameters as plasma density to produce REs. In total, 23 discharges provided useful data for this aggregate analysis, which focuses on the flattop $I_P$ phase ($t \approx$~0.5-1.5~s). Since the camera captures images at $\sim$60~fps, there are $\sim$1400 total time-slices, and synchrotron radiation was detected over the background plasma light and HXR speckles during $\sim$25\% of these times. At each time, the following RE-relevant parameters were evaluated:

\renewcommand{\theenumi}{\alph{enumi}}
\begin{enumerate}
    \item the electric field on-axis, calculated as $E_0 = V_{loop}/2\pi R_0$, using the external loop voltage measurement during $I_P \approx constant$;
    
    \item the theoretical Connor-Hastie threshold electric field \cite{connor1975}, $E_C= e^3 n \ln \Lambda /4 \pi \epsilon_0^2 m c^2 \propto n$;
    
    \item the Dreicer electric field \cite{dreicer1959,dreicer1960} required for thermal electrons to run away, $E_D = E_C \times mc^2/T \propto n/T$; 
    
    \item the characteristic synchrotron radiation timescale, $\tau_{rad} = 6 \pi \epsilon_0 m^3 c^3 /e^4 B^2 \propto 1/B^2$; and
    
    \item the RE collisional timescale, $\tau_{coll} = mc/e E_C \propto 1/n$.
\end{enumerate}

\noindent Here, $e$ is the electric charge, $m$ is the electron mass, and the Coloumb logarithm $\ln \Lambda = 15$ was assumed. From these parameters, three ratios are important: The first, $E_0/E_C \propto V_{loop}/n$, gives insight into the competition between the driving electric force and collisional friction on REs. The second, $E_0/E_D \propto V_{loop}\, T/n$, indicates the population of thermal electrons available to accelerate into the runaway regime. The third, $\tau_{rad}/\tau_{coll} \propto n/B^2$, compares the roles of synchrotron radiation damping and collisional drag on REs.

\begin{figure}[h]
    \centering
    \includegraphics[width=\textwidth]{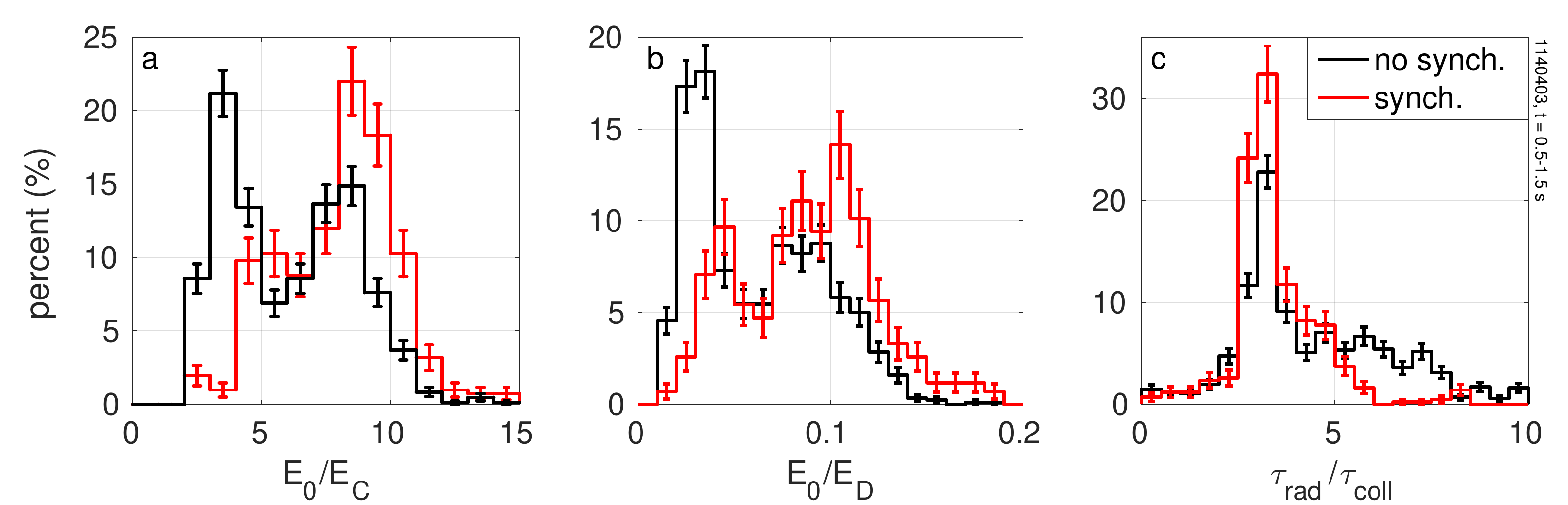}%{figures/histograms_EOverEc_EOverED_tRadHat_t0500-1500ms_1140403_180803.pdf}
    \caption{Histograms (with statistical error bars) of measured ratios of (a) on-axis to Connor-Hastie threshold electric fields, $E_0/E_C$ \cite{connor1975}; (b) on-axis to Dreicer electric fields, $E_0/E_D$ \cite{dreicer1959,dreicer1960}; and (c) characteristic radiation to collisional timescales, $\tau_{rad}/\tau_{coll}$, for flattop data when synchrotron radiation was (red) and was not (black) observed. ($t$~=~0.5-1.5~s)}
    \label{fig:histogram}
\end{figure}

Histograms in figure~\ref{fig:histogram} show the percentage of times during which synchrotron emission was (red) or was not (black) observed, binned for each ratio. Error bars are calculated as $\sqrt{N_{bin}}/N_{tot}$, where $N_{bin}$ is the number of counts in each bin, and $N_{tot}$ is the sum of counts in all bins. When synchrotron emission is observed, the distributions of $E_0/E_C$ and $E_0/E_D$ are shifted toward higher values (figure~\ref{fig:histogram}a-b), whereas that of $\tau_{rad}/\tau_{coll}$ is skewed toward lower values (figure~\ref{fig:histogram}c). This matches expectations, since higher $E_0/E_C$ and $E_0/E_D$ values lead to higher RE energies and densities; lower $\tau_{rad}/\tau_{coll}$ values, on the other hand, indicate more power lost --- and detected --- from synchrotron emission relative to that from collisions. 

It is important to note several subtleties associated with this analysis, especially since there is significant overlap in the histograms. First, spatial variation of RE parameters has been neglected, and only bulk plasma parameters were used. In addition, REs dynamically evolve in energy and number, so the RE population at the current time was affected by plasma parameters from earlier times. While time evolution has not been considered here, this analysis of aggregate data, collected at $\sim$60~Hz, still provides a general physical picture of the conditions under which REs will or will not produce detectable synchrotron emission. Due to the slow variation of plasma parameters during flattop $I_P$, multiple data points within this period can help counteract noise and improve statistics. Even then, the energy confinement time is $\sim$20-30~ms, similar to the time between camera frames, $\sim$17~ms.

%%%%%%%%%%%%%%%%%%%%%%
%%%%%%%%%%%%%%%%%%%%%%
% Spatiotemporal evolution
%%%%%%%%%%%%%%%%%%%%%%
%%%%%%%%%%%%%%%%%%%%%%

\section{Spatiotemporal evolution}\label{sec:spatiotemporal}

A wealth of information is stored in synchrotron images relating to the spatiotemporal evolution of the RE phase-space distribution. In this section, a detailed analysis of one C-Mod discharge (\#1140403026) is performed to infer the evolution of the radial density profile of the RE population. To completely solve the \emph{inverse} problem --- i.e. determine the 6D position and momentum-space distribution of REs from a 2D camera image --- is currently intractable. First, it is unlikely that there is a unique solution, especially within experimental uncertainties. Second, the computational resources required are expensive. Conversely, there is an opportunity to completely solve the \emph{forward} problem: Given 3D spatial distributions of all plasma parameters (e.g. $E$, $n$, $T$, $B$, etc.), one could solve the equations of motion for all electrons, calculate their synchrotron emission, and model its detection by a camera. Such a 6D solver, including the synthetic camera diagnostic, has been developed \cite{carbajal2017}; however, these simulations are computationally intensive, requiring hundreds of thousands of CPU hours. Instead, a much more computationally-feasible, multi-step approach is adopted in this paper to partly solve both the forward and inverse problems. The methodology is as follows:

\renewcommand{\theenumi}{\roman{enumi}}
\begin{enumerate}
    \item Flux-surface-averaged plasma parameters are obtained from experimental measurements for the magnetic axis and rational flux surfaces $q$~=~1, 4/3, 3/2, 2, and 3.
    
    \item For each surface, the RE momentum-space distribution function is evolved using the kinetic solver CODE \cite{landreman2014,stahl2016}. An ad hoc piecewise radial phase-space distribution, normalized to local RE density (see equation~(9) of \cite{landreman2014}), is constructed:
    
    \begin{samepage}
    \begin{eqnarray}
                      & F_{axis}\left(\vec{p}\right), \; & R \in \Big[ R_0, \left( R_0 + R_1 \right)/2 \Big) \nonumber \\
        F(R,\vec{p}) = \Bigg\{ \; & F_{q_m} \left(\vec{p}\right), & R \in \Big[ \left( R_{m-1} + R_m \right)/2, \left( R_m + R_{m+1} \right)/2 \Big) \label{eq:f} \\
                      & F_3 \left(\vec{p}\right), & R \in \Big[ \left( R_2 + R_3 \right)/2, R_0 + a \Big] \nonumber
    \end{eqnarray}
    \end{samepage}

    \noindent where $R$ is the major radial coordinate, $\vec{p} = (p_\parallel,p_\perp)$ is the 2D momentum vector, $q_m = \left\{1, 4/3, 3/2, 2 \right\}$ are the inner flux surfaces. Note that all $F_{q_m}$ and $R_m$ are also functions of time, but $t$ is dropped for convenience.
    
    \item This phase-space distribution, $F(R,\vec{p})$ from \eqref{eq:f}, is input into the synthetic diagnostic SOFT \cite{hoppe2018}, along with the magnetic topology and detector geometry, to generate a Green's function, $\hat{I}_{ij}\left(R\right)$, describing the partial image produced by the distribution function localized at radius $R$, as in equation (9) of \cite{hoppe2018}. Here, $i$ and $j$ refer to 2D pixel coordinates.
    
    \item \label{step:Iij} For an array of radial positions $R_k$, $\hat{I}_{ij}\left(R_k\right)$ is used as a set of basis functions, such that the final 2D image in the detector plane is
    
    \begin{equation}
        I_{ij} = \sum_k C(R_k) \, \hat{I}_{ij}(R_k) \, \Delta R_k,
        \label{eq:Iij}
    \end{equation}

    \noindent where $\Delta R_k$ is the radial step, and $C(R_k)$ can be calculated to produce the best fit between $I_{ij}$ and the experimental image. The RE density profile, $n_{RE}(R)$, can be related to coefficients $C(R_k)$ using $F(R,\vec{p})$, as described in \ref{app:A}.
    
\end{enumerate}

\subsection{Momentum-space simulations from CODE}\label{sec:CODE}

The kinetic Fokker-Planck solver, \emph{COllisional Distribution of Electrons} (CODE) \cite{landreman2014,stahl2016}, was used in this analysis to evolve RE momenta on each surface. Inputs to CODE are time evolutions of the electric field $E$, electron density $n$ and temperature $T$, toroidal magnetic field $B$, and effective charge $Z_{eff}$. In these experiments, a measurement of $Z_{eff}$ was unavailable as visible synchrotron light dominated the diagnostic measurement; thus, $Z_{eff}$~=~4 --- a value consistent with previous measurements during low density C-Mod discharges --- was assumed to be constant in time and space. The other parameters vary throughout the plasma: $n$ and $T$ radial profiles were measured with Thomson Scattering; $E$ and $q$ profiles were determined using EFIT \cite{lao1985}; and the toroidal magnetic field was approximated as $B = B_0 R_0/R$. The Chiu-Harvey knock-on collision model \cite{chiu1998} was used for avalanche generation, and all CODE simulations described here required $\sim$300 CPU hours in total.

\begin{figure}[h]
    \centering
    \includegraphics[width=\textwidth]{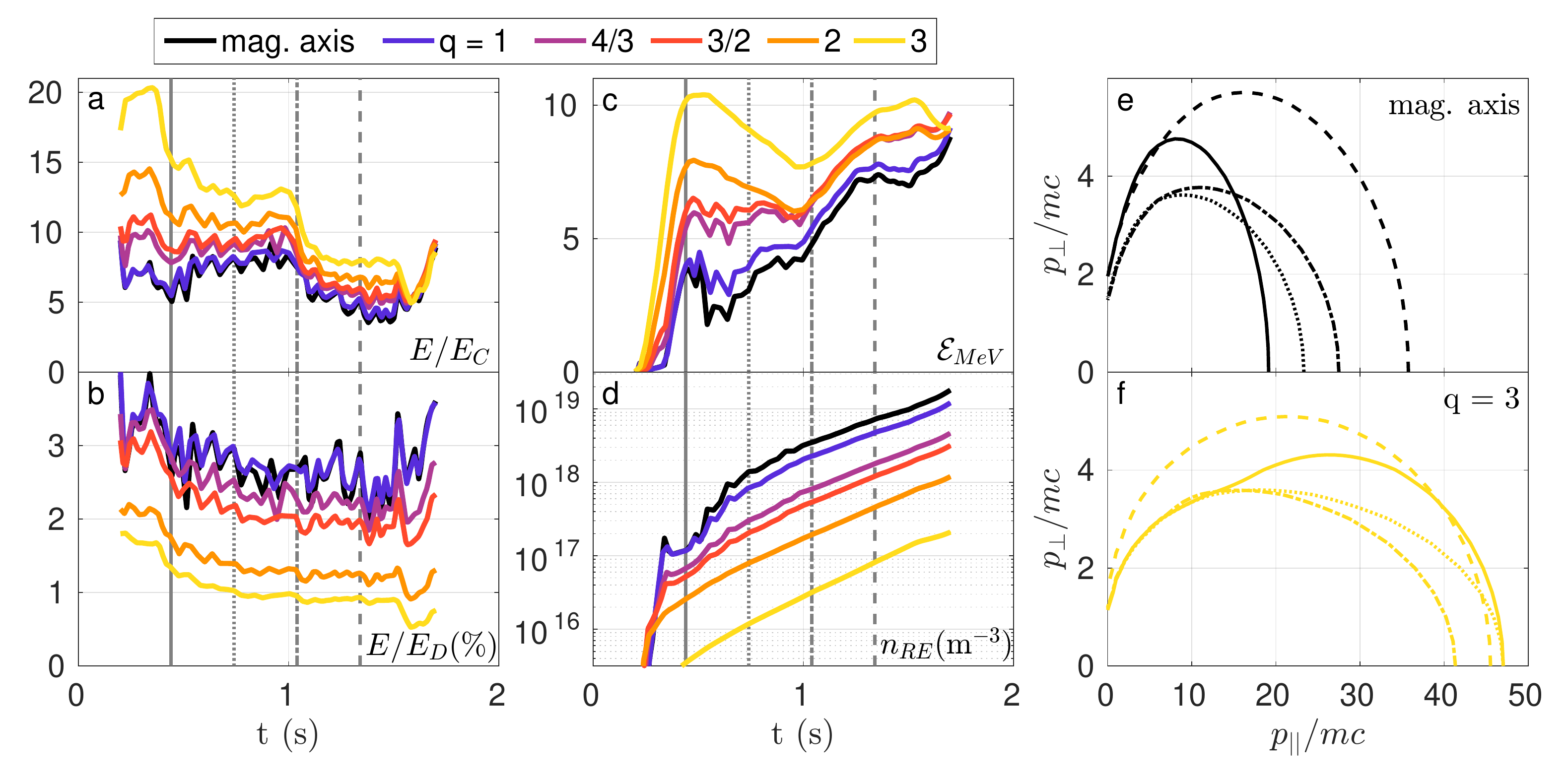}%{figures/180919/CODE_EOverEc_EOverED_EMeV_nRE_contours_maxis_q3_180919.pdf}
    \caption{Ratios (a) $E/E_C$ and (b) $E/E_D$ (\%) are provided as inputs to CODE \cite{landreman2014,stahl2016} for six radial positions: the magnetic axis and rational flux surfaces $q$~=~1, 4/3, 3/2, 2, and 3. CODE outputs the predicted (c) average RE energy (MeV) and (d) RE density (m$^{-3}$). Contours of CODE momentum-space distribution functions are shown for the (e) magnetic axis and (f) flux surface $q$~=~3 for four times, marked as vertical lines in (a)-(d): $t$~=~0.44 (solid), 0.74 (dotted), 1.04 (dot-dashed), and 1.34~s (dashed). Note that the value of each contour is (arbitrarily) chosen to be $\log_{10}(F)$~=~-15/4, where $F$ is normalized. The color scheme is the same as that in figure~\ref{fig:EFIT}. ($B_0$~=~5.4~T, $Z_{eff}$~=~4)}
    \label{fig:CODE}
\end{figure}

This analysis considers RE generation and evolution at six locations throughout the plasma: the magnetic axis and flux surfaces $q$~=~1, 4/3, 3/2, 2, and 3. These were chosen because they are approximately equally-spaced radially (see figure~\ref{fig:EFIT}) and, as rational surfaces, could potentially exhibit interesting behavior. Additionally, as will be described, all measurable synchrotron activity is found to occur within $q \leq 3$. Both $n$ and $T$ are assumed to be flux functions, but $E$ and $B$ were flux-surface averaged. The time evolutions of $E/E_C$ and $E/E_D$ are shown for each location of interest in figure~\ref{fig:CODE}a-b. Notice that $E/E_C$ increases radially from the magnetic axis to plasma edge; this is primarily due to the decreasing $n$ profile, but can also be affected by the radial dependence, $E \propto 1/R$, and finite diffusion time of changing $E$ into the plasma. From this, the RE average energy, $\mathcal{E}$, is expected to increase from the core to edge. Conversely, the values of $E/E_D$ are higher in the plasma center than at the boundary, due to the centrally-peaked $T$ profile. Thus, the runaway density, $n_{RE}$, is expected to be highest in the core as there is a larger population of thermal electrons available to run away. Figure~\ref{fig:CODE}c-d confirm these expectations: CODE predicts higher $\mathcal{E}$ on the surface $q=3$ compared to on-axis, whereas $n_{RE}$ is estimated to be approximately two orders of magnitude larger in the core than at the edge. 

The time evolution of $\mathcal{E}$ in figure~\ref{fig:CODE}c illustrates the complicated interplay of time-changing plasma parameters and RE dynamics. First, it is important to note that $\mathcal{E}$ is the \emph{average} energy of the high energy runaway region in momentum-space, not the maximum RE energy. Second, a finite time is required for REs to respond to a change in $E/E_C$ or $E/E_D$, meaning that $\mathcal{E}(t)$ will exhibit some time delay. In figure~\ref{fig:CODE}c, $\mathcal{E}$ increases rapidly during the $I_P$ ramp-up, but levels off or decreases as $E/E_C$ drops, even for values of $E/E_C \sim$~7-14. When the bulk plasma density increases at $t \approx$~1~s, $E/E_C$ decreases to $\sim$5-10, but $\mathcal{E}$ rises, due to both $E/E_C > 5$ and increased pitch angle scattering from increased collisionality. 

This can be seen from the contours of the normalized momentum-space distribution function at the magnetic axis and surface $q$~=~3 in figure~\ref{fig:CODE}e-f. The four times of interest are $t$~=~0.44 (solid), 0.74 (dotted), 1.04 (dot-dashed), and 1.34 s (dashed). On-axis, the distribution function, $F_{axis}$, initially has large pitch angles ($v_{\perp}/v_{\parallel} \approx 0.5$), but then elongates along $p_\parallel$ as REs are accelerated toroidally, along the direction of the magnetic field. From $t$~=~1.04-1.34~s, $F_{axis}$ grows in both $p_\parallel$ and $p_\perp$ with increased pitch angle scattering due to higher collisionality. On the surface $q$~=~3, the early distribution function, $F_3$, has high energies at early times due to high $E/E_C$ values, before decreasing in $p_\parallel$ and $p_\perp$. Similar to $F_{axis}$, $F_3$ spreads in $p_\perp$ later in time. While these are contours of the \emph{normalized} distribution function, the density of REs is actually predicted to increase exponentially from secondary avalanching, as seen in figure~\ref{fig:CODE}d.

\subsection{Synthetic images from SOFT}\label{sec:SOFT}

To best reproduce experimental synchrotron images, the synthetic diagnostic, \emph{Synchrotron-detecting Orbit Following Toolkit} (SOFT) \cite{hoppe2018}, was used for its synthetic camera capabilities. SOFT takes the following as inputs: the magnetic topology --- assumed to be axisymmetric --- obtained from EFIT \cite{lao1985}; detector specifications including geometry and spectral range (see table~\ref{tab:wide2}); and RE phase-space distribution, $F(R,\vec{p})$. The simulation initiates electrons with energies, pitches, and radial positions prescribed by $F(R,\vec{p})$ on the outer midplane ($R$~=~68-90~cm). The electrons follow their guiding center trajectories, conserving magnetic moment. If synchrotron radiation emitted by a RE is incident on the detector, its intensity and pixel location are recorded. As will be discussed, the use of a synthetic diagnostic, like SOFT, is of utmost importance in the analysis of synchrotron images as synchrotron radiation \emph{emitted} is not always \emph{detected}.

The full spectral and angular calculation of synchrotron emission is available in SOFT. However, because the angular spread of synchrotron emission is quite small ($\sim$1/$\gamma$, where $\gamma$ is the relativistic factor), a ``cone'' model --- where radiation is only emitted along the RE direction of motion --- serves as an adequate approximation of the full \emph{angular} formulation, as discussed in \cite{hoppe2018,hoppe2018d3d,tinguely2018}. Additionally, the cone model significantly reduces computation time, which for all SOFT simulations used in this work was $\sim$3700 CPU hours in total. Note that first-order corrections to the guiding center motion are not included in SOFT, meaning that drift orbits and associated effects have not been accounted for in this study. Using equation~(8) of \cite{tinguely2018}, the radial drift of a 20~MeV RE in a plasma with parabolic current density profile and $I_P$~=~800~kA is $r_d \leq$~3~cm. This is small but non-negligible for image analysis and should be investigated in future work.

\begin{figure}[h]
    \centering
    \includegraphics[width=0.6\textwidth]{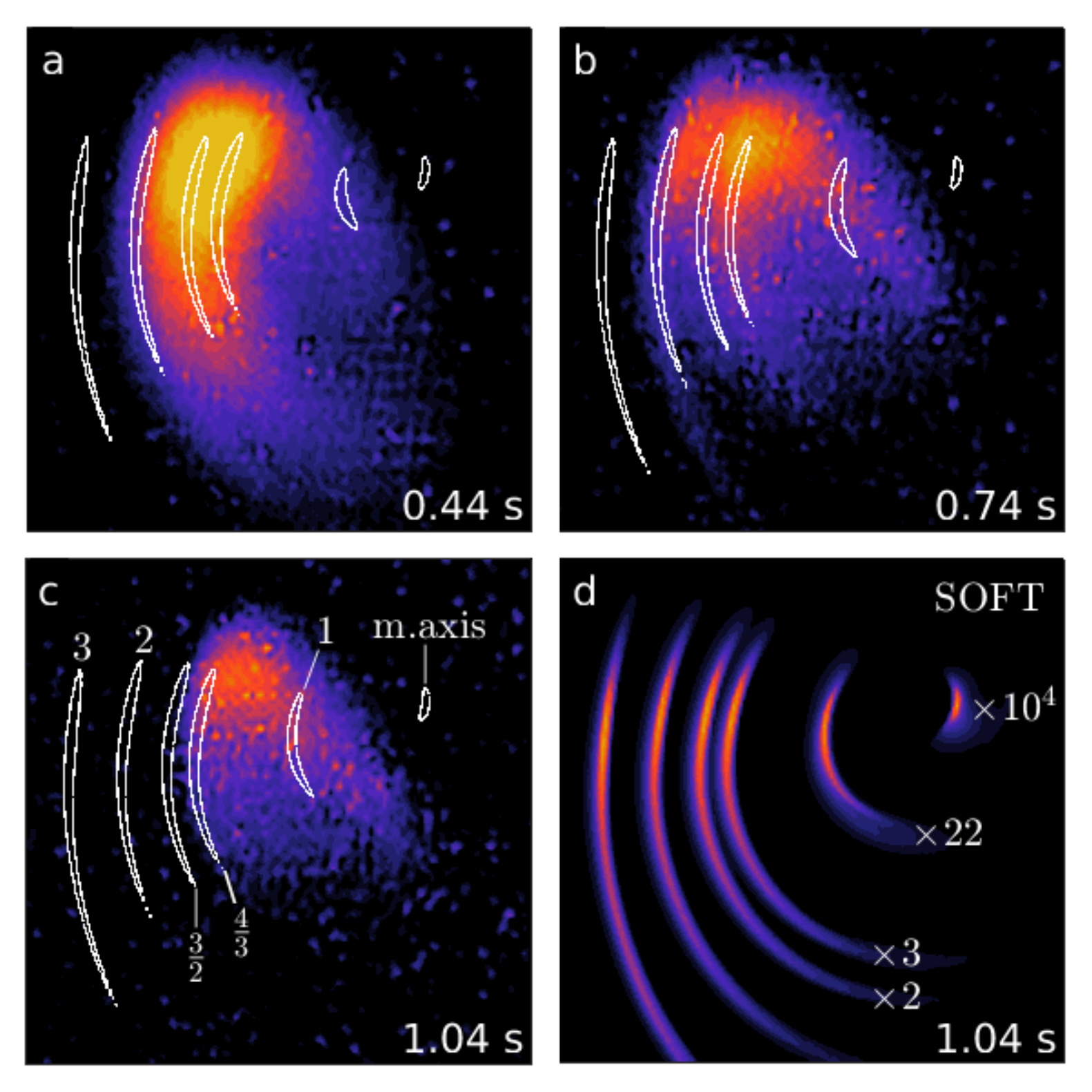}%{figures/SOFTContours_OverlayingWIDE2_1140403026_t0440-0740-1040ms_180803_v2.png}
    \caption{(a)-(c) Closed contours (white) of 50\% SOFT-predicted synchrotron emission \cite{hoppe2018}, from the magnetic axis and rational flux surfaces $q$~=~1, 4/3, 3/2, 2, and 3, overlay experimental images from figure \ref{fig:params}(a)-(c). (d) The full SOFT-predicted emission of each contour in (c) is shown, along with scaling factors required to plot all surfaces on the same color-scale. Note that the $q$-value increases from right-to-left in each subplot, as labeled in (c).}
    \label{fig:contours}
\end{figure}

One powerful feature of SOFT utilized in this work is its ability to calculate the Green's function, $\hat{I}_{ij}(R)$, which accounts for the momentum-space distribution as well as magnetic and detector geometries. Again, $i$ and $j$ are the 2D pixel coordinates. This function can be convolved with a radial density profile, $n_{RE}(R)$, as described in step~\ref{step:Iij}, to produce the final synchrotron image. Moreover, using $\hat{I}_{ij}(R)$, it is possible to identify the contribution of REs on a particular flux surface to the final image, simply by using a delta function at the flux surface location, i.e. $\delta (R-R_q)$. To highlight the contributions from REs near the magnetic axis and around flux surfaces $q$~=~1, 4/3, 3/2, 2, and 3, step functions of width 4~mm ($\Delta R/a \approx 2\%)$ centered on each surface were used instead of delta functions. For each surface, a closed contour at a level of 50\% maximum intensity, as predicted by SOFT, indicates the region of the image within which most of the synchrotron emission from REs on that surface will be detected. These contours are shown in figure~\ref{fig:contours}a-c overlaying the experimental images from figure~\ref{fig:params}a-c. As seen in each subplot, the contour grows in size and moves from right-to-left with increasing $q$-value (and $R$).

Even before considering the full intensity distribution predicted by SOFT, some spatial information can be gleaned from these images. Note how, in figure~\ref{fig:contours}a, the synchrotron spot shape matches the curvature of the SOFT contours quite well. In fact, the observed synchrotron spot is almost completely confined within the surface $q \sim 2$. This does not necessarily imply that REs exist only within $q \leq 2$; rather, REs at $q \geq 2$ could have too low energies ($<$~10~MeV) and/or densities to be detected by the camera. Figure~\ref{fig:contours}b shows the experimental image with interesting spatial structure at time $t$~=~0.74~s. Recall that at $t \approx$~0.7~s, increased MHD activity is observed at the onset of a locked mode. From the figure, it is clearly seen that the locations of surfaces $q$~=~3/2 and 2 match the inner and outer ``legs,'' respectively. If the locked mode is associated with a $m/n = 2/1$ tearing mode, then the camera may be capturing the radial transport of REs out of the plasma by an island at the rational surface $q$~=~2. At $t$~=~1.04~s, the synchrotron spot shrinks within the surface $q \approx$~4/3. This reduction in size could be due to (i) decreasing energies of REs located at $q \geq$~4/3 caused by the increasing plasma density and/or (ii) increased radial diffusion from the locked mode.

Figure~\ref{fig:contours}d is distinct from subplots \ref{fig:contours}a-c; here, the full SOFT-predicted synchrotron intensity distribution is shown for each SOFT contour in subplot \ref{fig:contours}c. The factors given next to the contours are those required to plot the intensities of magnetic axis and surfaces $q$~=~1, 4/3, and 3/2 on the same color-scale as surfaces $q$~=~2 and 3. The outermost surfaces are brightest due to the $\geq B^3$ scaling of synchrotron power spectra in the tokamak magnetic geometry \cite{hoppe2018d3d}.\footnote{The Larmor formula gives $P_{synch} \propto p_\perp^2 B^2$. Conservation of magnetic moment implies $p_\perp^2/B=constant$. Therefore, $P_{synch} \propto B^3$. This scaling can increase for specific wavelength ranges.} Geometric factors also affect the synchrotron radiation detected. Because the camera is far below the midplane, REs close to the magnetic axis and with small pitch angles --- which would otherwise dominate synchrotron emission in this scenario --- are not seen. The small contributions seen on-axis in figure~\ref{fig:contours}d come from particles with larger pitch angles, which are far less numerous. 

\subsection{Fit and reproduction of experimental images}\label{sec:results}

As seen in figure~\ref{fig:contours}, contributions from different flux surfaces to the final synchrotron image are almost non-overlapping. This is a consequence of the interplay between the high directionality of synchrotron emission and magnetic and detector geometries. Thus, one approach to reproduce the experimental images is to use $\hat{I}_{ij}(R_k)$ as a set of basis functions for discrete radial positions $R_k$. Then, coefficients $C(R_k)$, from \eqref{eq:Iij}, can be determined such that the resulting image, $I_{ij}$, best matches experimental data. Finally, the RE density profile, $n_{RE}(R)$, can be related to $C(R)$ through $F(R,\vec{p})$, as detailed in \ref{app:A}. The motivation for this fitting procedure is that while CODE has been used to construct a cylindrical plasma via $F(R,\vec{p})$, spatial dynamics --- such as drifts, diffusion, and trapping --- are not completely captured here. Therefore, fitting a density profile is actually \emph{required} to provide any useful spatial information about RE density evolution from the synchrotron images.

Note that equation~\eqref{eq:Iij} can be written as a matrix equation: Although the image, $I_{ij}$, is visualized in 2D, it can be represented as a 1D vector, $\mathbf{I}$, with $(i \times j)$ elements. Similarly, the coefficients, $C(R_k)$, also make up a 1D vector, $\mathbf{C}$, of length $k$, for the $k$ discrete values of $R$ used in the SOFT simulations. Then, the Green's function, $\hat{I}_{ij}$, can be rearranged into a 2D matrix, $\mathbf{\hat{I}}$, with dimensions $(i \times j) \times k$, so that

\begin{equation}
    \mathbf{I} = \mathbf{\hat{I}}\, \mathbf{C},
\end{equation}

\noindent where $\Delta R = constant$ has been absorbed. While the SOFT simulations were performed with REs at 200 radial positions, the output data were down-selected into $k$~=~100 radial bins to counteract over-fitting and best match pixel resolution. The best-fit coefficients were calculated using a linear least-squares solver subject to the constraint $\mathbf{C} \geq 0$, i.e. requiring non-negative contributions of synchrotron emission. In this analysis, values of $C(R)=0$ occurred in regions of low measured intensity near the plasma core and edge. This makes sense at the boundary of the synchrotron spot since we expect the RE radial profile to decay. However, values of $C(R) = 0$ in the plasma center, where high RE densities are expected, simply indicate our ignorance of the RE population there due to the camera's vertical offset.

\begin{figure}[h]
    \centering
    \includegraphics[width=\textwidth]{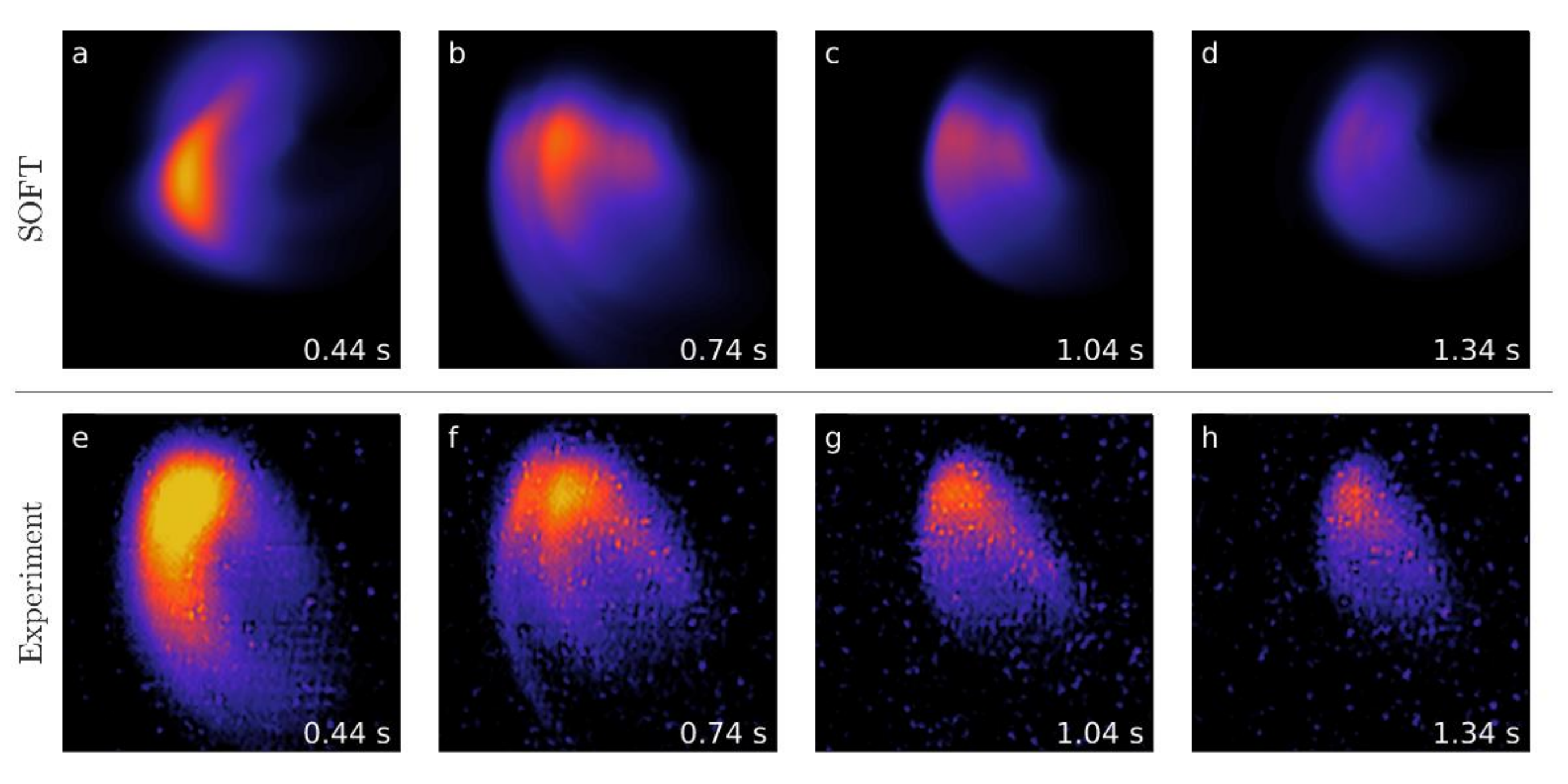}%{figures/180924/comparison_Exp_v_SOFT_allTimes_180924_v2.png}
    \caption{(a)-(d) Best-fit SOFT reproductions of the experimental images in subplots (e)-(h), reproduced from figure~\ref{fig:params}(a)-(d).}
    \label{fig:SOFT}
\end{figure}

Fitted images are shown for four times in figure~\ref{fig:SOFT}a-d, corresponding to the experimental images in figure~\ref{fig:SOFT}e-h. Note that each SOFT image is smoothed over a $5 \times 5$ pixel window ($\sim 3\%$) to remove unphysical sharp edges resulting from the piecewise structure of $F(R,\vec{p})$. For $t$~=~0.44~s, the fitted SOFT image does not match experiment well. This is likely because $I_P$ and the magnetic geometry are still evolving at this time. The simulated CODE momentum-space distributions also have large pitch angles at this time, causing more overlap in the contributions of adjacent flux surfaces to the final image. Additionally, the experimental image is saturated. For these reasons, a good fit of SOFT to the experimental image at $t$~=~0.44~s is difficult. The fitted images of later times, however, are more similar to experiment. As seen in figure~\ref{fig:SOFT}b, the spatial structure of the inner and outer ``legs'' can be reproduced by SOFT. The location of peak intensity and intensity gradients are also quite similar. Note that the vertical position of the synthetic image is slightly lower than that in experiment; a likely reason for this is calibration error, but the difference is $\leq$~10 pixels, which is small compared to the full size of the camera image ($640 \times 480$). The fitted SOFT images of later times, $t$~=~1.04 and 1.34 s in figure~\ref{fig:SOFT}c-d, also match experiment quite well, showing the decrease in both synchrotron spot size and intensity.  

\begin{figure}[h]
    \centering
    \includegraphics[width=\textwidth]{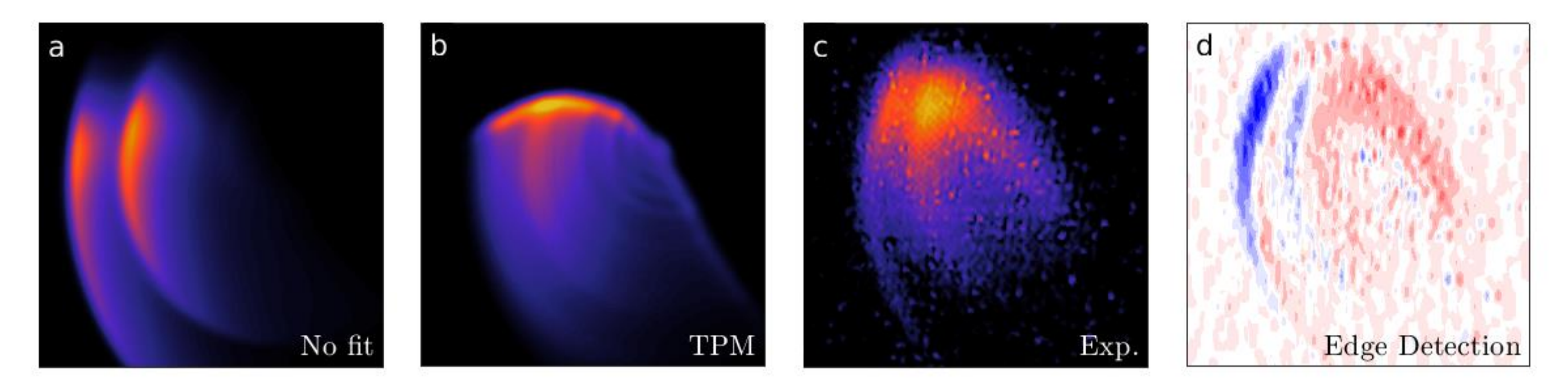}%{figures/180927/tpm_sobel_1140403026_180927_v1.png}
    \caption{(a) SOFT image resulting from no experimental fitting, but instead applying the CODE-predicted radial density profile (see figure~\ref{fig:CODE}d). (b) Best-fit SOFT reproduction using a test particle model of momentum-space evolution \cite{martin-solis1998}, described in section~\ref{sec:TPM}. (c) Experimental image from figure~\ref{fig:params}b reproduced for comparison. (d) Edge detection \cite{sobel1968} applied to (a), with blue/red colors corresponding to positive/negative horizontal gradients of pixel intensity, described in section~\ref{sec:edge}. ($t$~=~0.74~s)}
    \label{fig:tpm}
\end{figure}

As mentioned, there are modeling efforts currently underway to completely solve the RE forward problem, i.e. to accurately predict the 6D RE phase-space distribution from known plasma parameters. If the RE densities as calculated by CODE are used for the density profile, instead of the fitted profile, the resulting SOFT image at $t$~=~0.74~s would be that shown in figure~\ref{fig:tpm}a. Here, the two bright features correspond to the regions around the surfaces $q$~=~2 and 3. Specifically, the discontinuity in intensity occurs due to the piecewise nature of $F(R,\vec{p})$ and the assumed uniform density throughout each radial interval. Even though the RE densities are predicted to be $\sim$100$\times$ lower there than on-axis, geometric effects and increasing synchrotron power with magnetic field lead to CODE-predicted emission dominating near the edge. This synthetic image clearly does not match experiment, further motivating the fitting procedure employed in this work.

\subsection{Application of a test particle model}\label{sec:TPM}

In previous analyses, test particle models (TPMs) of RE evolution were used to estimate RE energies and pitches for comparison with experimental synchrotron images \cite{hollmann2015,yu2013,zhou2013,zhou2014,shi2010,chen2006,tong2016,finken1990,jaspers1994}. In this analysis, the TPM from \cite{martin-solis1998} was applied and found to be insufficient in capturing all spatial features of the experimental intensity distributions. In \cite{martin-solis1998}, a coupled system of differential equations describes the trajectory of a test RE in momentum-space, governed by time-evolving plasma parameters. These trajectories (delta functions in momentum-space) were calculated for 40 equally-spaced flux surfaces throughout the plasma, resulting in radial profiles of RE energy and pitch. Following a similar fitting procedure as that described in section~\ref{sec:results}, the best-fit TPM+SOFT image for $t$~=~0.74~s is shown in figure~\ref{fig:tpm}b. The fitted TPM image is still able to match the horizontal width of the synchrotron spot, and the intensity distribution is similar to experiment, e.g. the parabolic shape and approximate location of maximum intensity. However, the vertical extent of the synchrotron spot is not reproduced by the TPM (compare to figure~\ref{fig:tpm}c). This is simply because there is no pitch angle distribution, leading to a sharp intensity gradient at the top of the image. Thus, it is concluded that the full momentum-space distribution is needed to reproduce the smooth intensity gradients at the edges of the synchrotron spot.

\subsection{Edge detection}\label{sec:edge}

The full analysis of spatial intensity distributions within synchrotron images can be time consuming and computationally expensive. Yet useful spatial information can still be obtained from SOFT without knowing the full momentum-space evolution. Specifically, spatial structure in the synchrotron image can be mapped to flux surfaces in the plasma, assuming that most observed REs have small pitch angles ($v_{\perp}/v_{\parallel} \leq$~0.2). To demonstrate this, the Sobel operator \cite{sobel1968} for edge detection was applied to the experimental data. Full details of the calculation are included in \ref{app:B}, but the general idea is that the Sobel operator approximates the gradient of pixel intensity within the image. Here, the \emph{horizontal} gradient operator was used to identify \emph{vertical} edges in the images. 

The resulting horizontal gradient of pixel intensity in the experimental image, at $t$~=~0.74~s, is shown in figure~\ref{fig:tpm}d. The blue/red colormap corresponds to the amplitude of positive/negative gradients of pixel intensity in the horizontal direction, from left-to-right. Thus, the left edge of the synchrotron spot is blue, and right edge is red. By setting a threshold in gradient, the edges can be detected, and the pixel locations can be mapped to flux surfaces using SOFT. Notice that in the frame shown, there are two blue regions, indicating that there is more complicated spatial structure. These are the inner and outer  ``legs'' of the synchrotron spot. The white region (zero slope) between blue and red (positive and negative slopes, respectively) were used to identify the spatiotemporal evolution of these legs, which will be described in the next section.

\subsection{RE density profile}\label{sec:density}

\begin{figure}[h]
    \centering
    \includegraphics[width=0.65\textwidth]{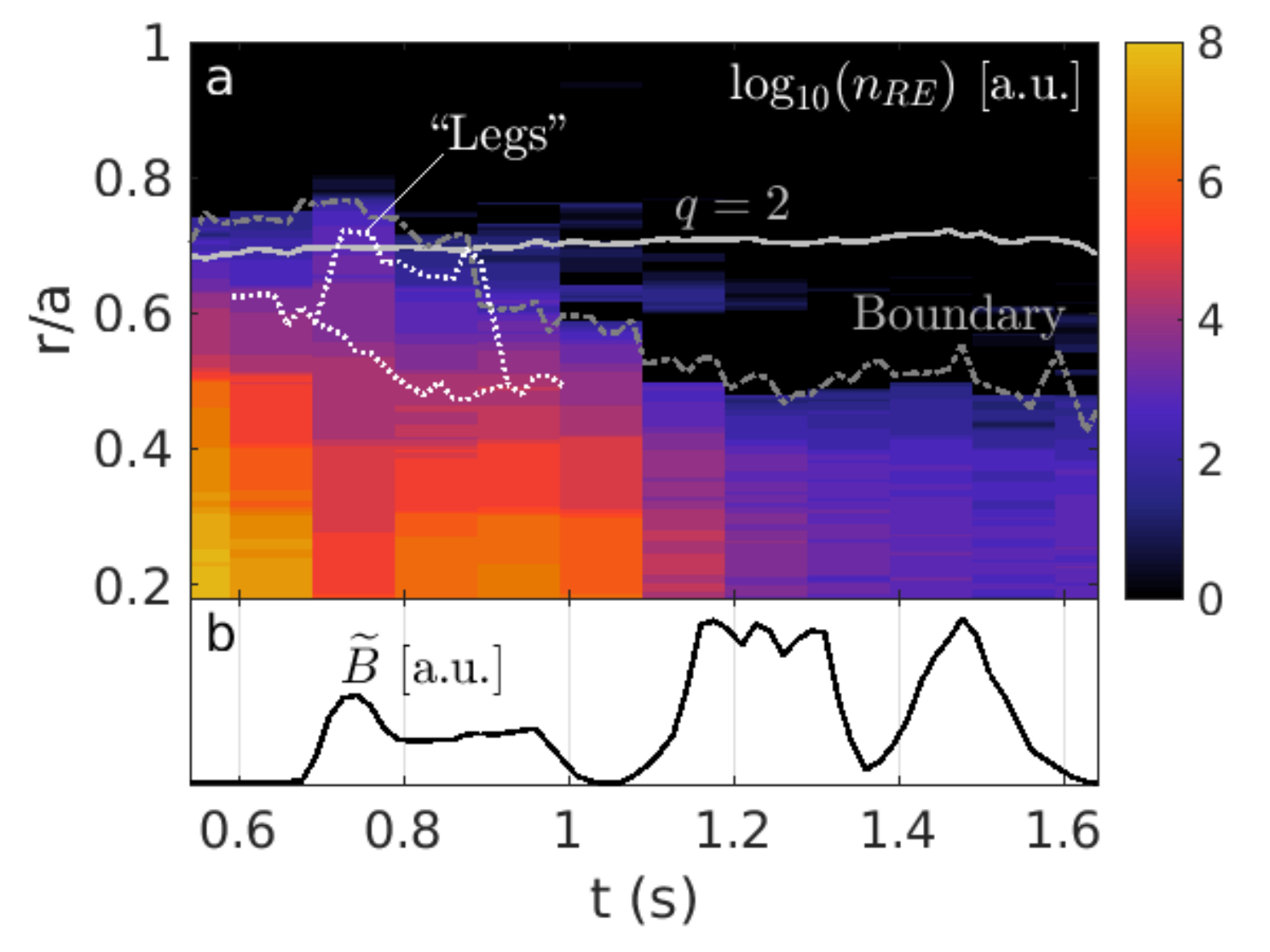}%{figures/180919/nRE_v_t_rOvera_wMagneticSignal_180919.png}
    \caption{(a) Contour plot of best-fit RE density (a.u.) versus time and normalized minor radius, $r/a$. The boundary (dot-dashed) and ``legs'' (dotted), as determined by edge detection, as well as the surface $q=2$ (solid), overlay the radial profile. Time and radial resolutions are $\Delta t \approx$~100~ms and $\Delta r/a \approx 0.01$, respectively. (b) Reproduction of the magnetic fluctuation signal (a.u.) from figure~\ref{fig:params}e.}
    \label{fig:nRE}
\end{figure}

\begin{figure}[h]
    \centering
    \includegraphics[width=0.65\textwidth]{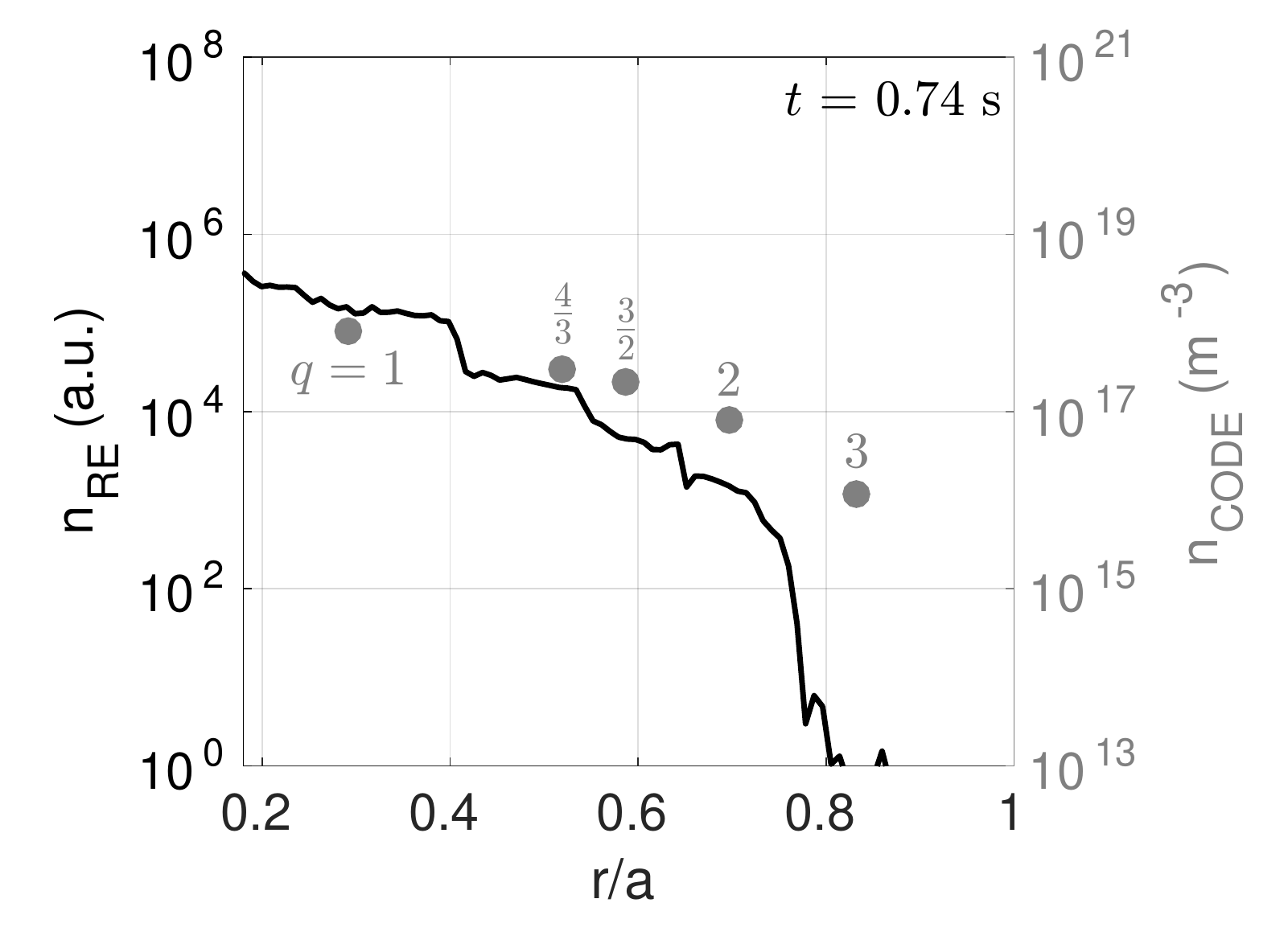}%{figures/180919/nRE_Fit_CODE_v_rOvera_t0740ms_180919.pdf}
    \caption{(Left axis) The fitted $n_{RE}$ radial profile (solid) at time $t$~=~0.74~s, plotted on the same radial and logarithmic scales as figure~\ref{fig:nRE}a. (Right axis) The CODE-predicted RE densities, $n_{\mathrm{CODE}}$ (dots), for the surfaces in figure~\ref{fig:CODE}d, also at $t$~=~0.74~s and spanning 8 orders of magnitude.}
    \label{fig:profile}
\end{figure}

The resulting best-fit RE density profile is plotted versus time and normalized minor radius, $r/a$, in figure~\ref{fig:nRE}a, with one time, $t$~=~0.74~s, highlighted in figure~\ref{fig:profile}. Only times during flattop $I_P$ are shown as the fitting procedure did not reproduce experimental images before $t \approx$~0.5~s (see figure~\ref{fig:SOFT}a). In addition, the spatial range only spans $r/a \sim$~0.2-1.0, since little synchrotron emission is visible from REs near the magnetic axis by a camera displaced below the midplane; this makes interpretation of results for $r/a \leq 0.2$ difficult. (For reference, the surface $q=1$ is located at $r/a \approx 0.3$.) The time resolution of this analysis is $\Delta t$~=~100~ms, limited primarily by the long computation times of SOFT simulations of each time-slice. While the radial resolution of the SOFT simulations is $\Delta r/a \approx 0.01$, caution should be used when interpreting fine spatial features. Note also that the colormap of $\log_{10}(n_{RE})$ has arbitrary units since no absolute calibration of the camera was performed. It follows that the density threshold for detection cannot be inferred from this analysis; here, the scale spans 8 orders of magnitude, which adequately portrays the spatiotemporal evolution while also highlighting interesting spatial features.

As seen in figures~\ref{fig:nRE}a and \ref{fig:profile}, the radial profile is peaked in the core and decays toward the plasma edge. In figure~\ref{fig:nRE}a, at the start of the flattop, the observed RE beam is confined within $r/a \leq 0.75$. At $t \approx$~0.7~s, $n_{RE}$ spreads outward toward the surface $q$~=~2, corresponding to the start of MHD activity (reproduced in figure~\ref{fig:nRE}b) and spatial structure observed at that time. Figure~\ref{fig:profile} shows the radial profile for $t$~=~0.74~s. Note that the profile decreases (approximately) monotonically; i.e. there is no significant ``bump'' in $n_{RE}$ near the surface $q=2$ ($r/a \approx 0.7$). There are, however, several ``steps'' around $r/a \approx$~0.4, 0.55, and 0.65, which are caused by the piecewise momentum-space distribution in \eqref{eq:f}. Beyond $r/a \approx 0.75$, $n_{RE}$ drops off steeply. This is in contrast with the RE density ``profile'' calculated by CODE, also shown in figure~\ref{fig:profile}, which is broader in radial extent and predicts far higher densities at $q$~=~3 than inferred. For times after $t \approx$~0.74~s, $n_{RE}$ shrinks in size and amplitude, as seen in figure~\ref{fig:nRE}a. This is consistent with the bulk plasma density increasing, thereby suppressing the RE population.

Overlaying the contour plot of figure~\ref{fig:nRE}a are time evolutions of the plasma boundary and ``legs'' as determined by edge detection (see section~\ref{sec:edge}), as well as the location of the surface $q=2$. The boundary (dot-dashed) matches the shape of the density profile quite well. The low density ``bumps'' outside this boundary are likely due to mis-identification of some HXRs --- with low enough intensity to be missed by the data filtering technique --- as synchrotron light. The inner and outer legs (where the dotted lines diverge) form at the same time that magnetic fluctuations and a locked mode are observed, as seen in figure~\ref{fig:nRE}b. From $t \approx$~0.7-0.9~s, the inner leg follows the density contour ($\log_{10}(n_{RE}) \approx 5$ in figure~\ref{fig:nRE}a), while the outer leg moves to region near the surface $q \approx$~2 (solid). At $t \approx$~0.9~s, the two legs recombine.

%%%%%%%%%%%%%%%%%%%%%%
%%%%%%%%%%%%%%%%%%%%%%
% Summary
%%%%%%%%%%%%%%%%%%%%%%
%%%%%%%%%%%%%%%%%%%%%%

\section{Summary}\label{sec:summary}

In this work, relativistic runaway electrons (REs) were studied during the flattop $I_P$ phase of low density, diverted plasma discharges in Alcator C-Mod. Because of C-Mod's high toroidal magnetic field ($B_0$~=~5.4~T), high energy REs emit synchrotron radiation in the visible wavelength range ($\lambda \approx$~400-900~nm); consequently, images of this emission were captured by a wide-angle camera viewing both co- and counter-$I_P$ directions inside the tokamak. An in-vessel calibration was performed to correct for image distortion, thereby allowing the diagnosis of ``in-flight'' RE spatiotemporal evolution. Furthermore, the study of synchrotron images is motivated by diagnostic opportunities of future fusion devices, which will have the potential to measure visible synchrotron radiation due to their high magnetic fields; e.g. on ITER ($\sim$5~T), SPARC ($\sim$12~T) \cite{greenwald2018}, and ARC ($\sim$9~T) \cite{sorbom2015}.

A statistical analysis of aggregate data from 23 RE-producing discharges ($>$1000 camera images) explored the plasma parameter space for regions in which RE synchrotron radiation was or was not detected by the visible camera. In general, visible synchrotron emission was observed for higher values of $E_0/E_C$ \cite{connor1975} and $E_0/E_D$ \cite{dreicer1959,dreicer1960} and lower values of $\tau_{rad}/\tau_{coll}$ compared to the times when synchrotron emission was \emph{not} observed. This matches theoretical predictions: Higher $E_0/E_C$ and $E_0/E_D$ lead to higher RE energies and larger RE growth rates, thus increasing the likelihood of detection of visible synchrotron light. In addition, low $\tau_{rad}/\tau_{coll}$ ($<10$) values typically indicate that synchrotron radiation is dominating collisional friction as a power loss mechanism.

For one discharge, the spatiotemporal evolution of REs was explored in detail through the analysis of synchrotron image evolution. Both a test particle model (TPM) \cite{martin-solis1998} and kinetic solver CODE \cite{landreman2014,stahl2016} were used to simulate RE dynamics in momentum-space at many positions throughout the plasma, including the magnetic axis and rational flux surfaces $q$~=~1, 4/3, 3/2, 2, and 3. The resulting distributions of energy and pitch angle were input into the synthetic camera diagnostic SOFT \cite{hoppe2018}, which also accounts for geometric effects of the magnetic topology and detector specifications. Using SOFT to generate a Green's function allowed identification of contributions from each flux surface to the final synthetic image. Therefore, edges detected \cite{sobel1968} in the experimental images were mapped to the flux surface ``boundary'' of the synchrotron spot; this time-evolving boundary was observed to decrease in size with increases in both plasma density and MHD activity. In addition, an interesting spatial feature was measured at the onset of a locked mode; a third ``leg'' of the synchrotron spot was found to be located approximately at the rational surface $q$~=~2. Such spatial structure could indicate that REs were trapped in a 2/1 island and expelled from the plasma due to increased radial transport.

Moving beyond the identification of spatial features only, the experimental synchrotron intensity \emph{distribution} within the images was also investigated. Due to the non-overlapping nature of SOFT-predicted emission from different flux surfaces, the Green's function was utilized as a set of basis functions, from which a synthetic image could be constructed. In this way, the SOFT synthetic synchrotron images were fit to experiment, producing a RE density profile evolution. Fitted synthetic images were found to match experiment well during the flattop $I_P$ phase, but those during the $I_P$ ramp-up did not. In addition, it was seen that the TPM could not reproduce all spatial features of experimental images; therefore, the full momentum-space distributions from CODE were needed for a complete analysis. Such a procedure as that adopted in this paper could be used to study the spatiotemporal dynamics of REs in other current and future tokamaks. Furthermore, SOFT's capabilities should be utilized in this way to design --- and generate synthetic data for --- new and better RE diagnostic setups. 

%%%%%%%%%%%%%%%%%%%%%%
%%%%%%%%%%%%%%%%%%%%%%
% Acknowledgements
%%%%%%%%%%%%%%%%%%%%%%
%%%%%%%%%%%%%%%%%%%%%%

\section*{Acknowledgements}

The authors thank J. Terry, R. Mumgaard, S. Scott, J. Hughes, and A. Patterson for fruitful discussions, as well as the entire Alcator C-Mod team. This work was supported by US DOE Grant DE-FC02-99ER54512, using Alcator C-Mod, a DOE Office of Science User Facility; Vetenskapsr\aa det (Dnr 2014-5510); and the European Research Council (ERC-2014-CoG grant 647121).

%%%%%%%%%%%%%%%%%%%%%%
%%%%%%%%%%%%%%%%%%%%%%
% Appendices
%%%%%%%%%%%%%%%%%%%%%%
%%%%%%%%%%%%%%%%%%%%%%

%%%%%%%%%%%%%%%%%%%%%%
% Appendix A
%%%%%%%%%%%%%%%%%%%%%%

\appendix
\section{Calculation of the RE density profile}\label{app:A}

The output momentum-space distribution function $F(\vec{p})$ from CODE is normalized, as given by equation~(9) of \cite{landreman2014}:

\begin{equation}
    F(\vec{p}) = \frac{\pi^{3/2}\, m^{3}\, v_{ref}^3}{n_{ref}} f(\vec{p}),
\end{equation}

\noindent where $m$ is the electron mass, $v_{ref}$ is a reference electron thermal velocity, $n_{ref}$ is a reference electron density, and $\int f(\vec{p}) \, \mathrm{d}\vec{p} = n$, the total plasma density. Thus, analagous to \eqref{eq:f}, the \emph{unnormalized} ad hoc profile would be

\begin{equation}
    f(R,\vec{p}) = \frac{n_{ref}(R)}{\pi^{3/2}\, m^3\, v_{ref}^3(R)} F(R,\vec{p}).
\end{equation}

\noindent Here, $n_{ref}(R)$ and $v_{ref}(R)$ are also piecewise functions like $F(R,\vec{p})$. CODE also calculates the total RE density, $n_{\mathrm{CODE}}$, like that plotted in figure~\ref{fig:CODE}d. In this work, the Green's function, $\hat{I}_{ij}(R)$, from \eqref{eq:Iij}, was calculated using normalized $F(R,\vec{p})$ instead of $f(R,\vec{p})$. Therefore, the fitted RE density profile is calculated

\begin{equation}
    n_{RE}(R) = \frac{\pi^{3/2}\, m^{3}\, v_{ref}^3(R)}{n_{ref}(R)} n_{\mathrm{CODE}}(R) \, C(R), 
\end{equation}

\noindent where $C(R)$ are the best-fit coefficients of \eqref{eq:Iij}. Note that all quantities that vary with $R$ also vary in time, but $t$ has been dropped for clarity.

%%%%%%%%%%%%%%%%%%%%%%
% Appendix B
%%%%%%%%%%%%%%%%%%%%%%
\section{Calculation for edge detection}\label{app:B}

\noindent While many feature detection algorithms exist for image processing, the edge detection analysis performed in this work (see section~\ref{sec:edge}) uses the Sobel operator \cite{sobel1968},

\begin{eqnarray}
                            & -1 \;\;\;\;\;\; 0 \;\;\;\;\;\; 1 & \nonumber \\
    \mathbf{S} = \Bigg[ \;  & -2 \;\;\;\;\;\; 0 \;\;\;\;\;\; 2 & \;\;\; \Bigg] \; , \label{eq:S} \\
                            & -1 \;\;\;\;\;\; 0 \;\;\;\;\;\; 1 & \nonumber
\end{eqnarray}

\noindent which is simple and efficient in implementation. Here, $\mathbf{S}$ is written in the form for calculation of the (approximate) horizontal gradient of pixel intensities when convolved with an image. This matrix can be rotated to calculate the vertical gradient or have signs flipped to switch gradient directions. 

Consider a pixel at location $(i,j)$ in an image, as well as its neighboring pixels. A subset of the total image, $I_{ij}$, is the $3 \times 3$ matrix surrounding $(i,j)$:

\begin{eqnarray}
                            & I_{i-1,j-1} \;\;\; I_{i,j-1} \;\;\; I_{i+1,j-1} & \nonumber \\
    \mathbf{\widetilde{I}_{ij}} = \Bigg[ \;\;  & I_{i-1,j} \;\;\;\;\;\; I_{i,j} \;\;\;\;\;\; I_{i+1,j} & \;\; \Bigg] \; . \label{eq:I3} \\
                            & I_{i-1,j+1} \;\;\; I_{i,j+1} \;\;\; I_{i+1,j+1} & \nonumber
\end{eqnarray}

\noindent From \eqref{eq:S} and \eqref{eq:I3}, the resulting ``gradient image,'' $\mathbf{I_{ij}'}$, used for edge detection is evaluated by (i) element-wise multiplication of $\mathbf{S}$ and $\mathbf{\widetilde{I}_{ij}}$ and (ii) summation over all (nine) elements. Explicitly, the value of the Sobel horizontal gradient at pixel location $(i,j)$ is

\begin{equation}
    I_{ij}' = -I_{i-1,j-1} + I_{i+1,j-1} - 2 I_{i-1,j} + 2 I_{i+1,j} - I_{i-1,j+1} + I_{i+1,j+1} \; .
\end{equation}

\noindent Note that this calculation cannot be performed at the edges of the image.

%%%%%%%%%%%%%%%%%%%%%%
%%%%%%%%%%%%%%%%%%%%%%
% References
%%%%%%%%%%%%%%%%%%%%%%
%%%%%%%%%%%%%%%%%%%%%%

\section*{References}
\bibliographystyle{unsrt}
\bibliography{bib}

\end{document}